\documentclass[twocolumn,secnumarabic,amssymb, nobibnotes, aps, prl, superscriptaddress, nobalancelastpage,longbibliography,nofootinbib]{revtex4-1}

 \usepackage{graphicx}
\usepackage{bm}
\usepackage{amsmath} \usepackage{braket} \usepackage{epsfig}
\usepackage{tensor}
\usepackage[version=4]{mhchem}
\usepackage{titlesec}
\usepackage{ifthen}
\usepackage{sidecap}
\usepackage{listings}
\usepackage[para,online,flushleft]{threeparttablex}

\usepackage{xr-hyper}
\usepackage{hyperref}
\hypersetup{breaklinks=true,colorlinks=true,linkcolor=blue,citecolor=blue,filecolor=magenta,urlcolor=cyan}

\usepackage[all]{hypcap}

\usepackage{ulem}

\usepackage{xcolor}
\definecolor{pastelgray}{rgb}{0.81, 0.81, 0.77}
\definecolor{beaublue}{rgb}{0.9, 0.9, 0.93}

\makeatletter
\def\@bibdataout@aps{%
\immediate\write\@bibdataout{%
@CONTROL{%
apsrev41Control%
\longbibliography@sw{%
    ,author="08",editor="1",pages="1",title="0",year="1"%
    }{%
    ,author="08",editor="1",pages="1",title="",year="1"%
    }%
  }%
}%
\if@filesw \immediate \write \@auxout {\string \citation {apsrev41Control}}\fi
}
\makeatother

\newcommand{\aD}{\alpha_{\rm D}}
\newcommand{\skin}{R_{\rm skin}}
\newcommand{\SV}{SV-min}
\newcommand{\Fy}{Fy($\Delta r$, HFB)}
\newcommand{\Rmir}{\Delta R_{\rm ch}^{\rm mir}}
\newcommand{\Rch}{R_{\rm ch}}

\renewcommand{\vec}[1]{\mbox{\boldmath $#1$}}
\newcommand{\vecs}[1]{\mbox{\boldmath \scriptsize$#1$}}

\begin{document}

\title{Information content of the differences in the charge radii of mirror nuclei}

\author{Paul-Gerhard Reinhard}
\affiliation{Institut für Theoretische Physik, Universität Erlangen, Erlangen, Germany}


\author{Witold Nazarewicz}
\affiliation{Facility for Rare Isotope Beams and Department of Physics and Astronomy, Michigan State University, East Lansing, Michigan 48824, USA}

\date{\today}
\begin{abstract}
Differences in the charge radii of mirror nuclei have been recently suggested to  contain information on the slope of the symmetry energy $L$. To test this hypothesis, we perform statistical correlation analysis 
using quantified  energy density functionals that are consistent with our previous knowledge on  global nuclear observables such as binding energies and charge radii.
We conclude that the difference in charge radii between a mirror pair, $\Rmir$, is an inferior isovector indicator compared to other observables, such at the neutron skin or electric dipole polarizability $\aD$. In particular, this quantity correlates poorly with both the neutron skin and $L$. We  demonstrate that $\Rmir$ is  influenced by pairing correlations in the presence of low-lying proton continuum in the proton-rich mirror-partner nucleus. Considering the  large theoretical uncertainties on $\Rmir$, we conclude that the precise data on mirror charge radii cannot provide a stringent constraint on $L$.
\end{abstract}
\maketitle

{\it Introduction}.---A cursory information on  properties of atomic nuclei is offered by the droplet model \cite{Wei35a,Mye77aB} whose key parameters are the nuclear matter (or
bulk) characteristics such as volume energy $E/A$, equilibrium density
$\rho_\mathrm{eq}$, incompressibility $K$, symmetry energy $J$, and
 symmetry energy  slope $L$. These quantities
are widely
used to characterize and compare nuclear models. The isoscalar
parameters ($E/A$, $\rho_\mathrm{eq}$, $K$) are well determined by
empirical data because the chart of nuclei extends over a large range
of mass numbers. However, the isovector parameters $J$ and $L$ are poorly
constrained because available isotopic chains are fairly short. This is an
uncomfortable situation because extrapolations to very neutron rich
isotopes and to neutron stars are crucial in nuclear astrophysics
\cite{Oezel2016,Oertel2017,Roca-Maza2018,Lattimer2021}. Consequently, there is  a
great demand for  isovector-sensitive data that can be used for constraining the symmetry energy in the various nuclear models.
The most promising  isovector indicators \cite{Reinhard2010} include neutron radii, neutron skins, dipole polarizability, and parity-violating asymmetry, see, e.g.,
Refs.~\cite{Brown2000,Horowitz20001,Furnstahl2002,Centelles2009,Roca-Maza2011,Piekarewicz2012,Nazarewicz2014,Essick2021,ReinhardPREX}.

Recently, it has been suggested~\cite{Brown2017}
that a difference in the charge radii of mirror nuclei, $\Rmir({^A}X/Y)\equiv  \Rch({^A_Z}X_N)- \Rch({^A_N}Y_Z)$, can serve as an isovector indicator that can be used to estimate the symmetry energy parameter $L$. Such estimates can be found in Refs.~\cite{Brown2017,Yang2018,Brown2020,Pineda2021,Novario2021} for the $^{50}$Ni/Ti, $^{52}$Ni/Cr, $^{54}$Ni/Fe, and $^{36}$Ca/S  mirror pairs.
It was concluded \cite{Brown2017,Yang2018} that the precise data on mirror charge radii can provide a stringent constraint on $L$. In this study, we examine this finding.

Since the mirror nuclei considered in the previous work are all open-shell systems, pairing correlations, ignored in Refs. ~\cite{Brown2017,Yang2018}, are expected to play a role in the analysis of  $\Rmir$. 
In this context, we first note that the proton-rich mirror partners considered in  Refs.~\cite{Brown2017,Yang2018,Brown2020,Pineda2021} are  all weakly bound, which appreciably affects pairing correlations due to the pair scattering into the continuum space. 
For such nuclei, to avoid the appearance of an unphysical particle gas that can impact radial behavior of nucleonic densities, nucleonic pairing must be handled within the full Hartree-Fock-Bogoliubov (HFB) scheme instead of the simpler Bardeen-Cooper-Schrieffer (BCS) approximation \cite{Dobaczewski1984,Dobaczewski1996,Dobaczewski2013}. Recently, this has been discussed in relation to the $^{36}$Ca charge radius measurement \cite{Mil19}. 
We also note that the effective pairing interaction in atomic nuclei has a strong isovector component, that is, a larger strength is required in the proton pairing channel than in the neutron pairing channel \cite{Bertsch2009}, and this can impact $\Rmir$. 

To reduce theoretical uncertainties,  a better understanding of model-dependent relationships between symmetry-energy parameters and nuclear observables is required. This can be achieved by means of the statistical correlation analysis \cite{Reinhard2010,Fattoyev2011} based on covariances obtained during the model calibration. Since the number of  conceivable observables of interest is enormous, and the number of model parameters is significant (usually greater than 10), 
the  covariance analysis is the least biased and most exhausting way to find out the correlations  between all conceivable observables as it explores the whole parameter space of a given model. 

In this work, we apply the HFB framework to assess the impact of weak binding and pairing on $\Rmir$. To study how strong and meaningful is the suggested correlation between $\Rmir$ and $L$, we use the  statistical covariance framework. 

{\it Theoretical models}.---Our analysis has been carried out with non-relativistic nuclear density-functional theory in its self-consistent  nuclear energy density functional (EDF) variant \cite{Bender2003}. 
In our applications, we employ the Skyrme parametrizations {\SV}  \cite{Klupfel2009}
and {\Fy} \cite{Rei17a,Mil19},
which have been optimized to a large experimental calibration datasets  by means of the standard linear regression, which provides
information on  uncertainties and statistical correlations
between observables.

Because we are dealing with weakly bound nuclei, pairing is treated at
the HFB level using an iterative scheme in
terms of canonical orbitals \cite{Rei97a,Tajima2004,Chen2021}.  The
pairing space is limited by a soft cutoff in single-particle space
with a Woods-Saxon profile \cite{Kri90a,Rei21aR,Chen2021}.  The pairing window involves
canonical states in a band of 15\,MeV above the Fermi level for Fy($\Delta
r$,HFB) and 5 MeV for SV-min(HFB)  with a smoothing of a tenth of the
of the pairing band width.   The pairing window of SV-min(HFB) is chosen to
comply with that of SV-min in the BCS approximation \cite{Klupfel2009}. The EDF SV-min(HFB) has been calibrated
using exactly the same strategy and dataset as for SV-min
\cite{Klupfel2009}. The changes in the model parameters when going from BCS to HFB are  small,  nonetheless important to
maintain the high quality of the parametrization. 
The HFB calculations discussed in this study are compared to BCS results and to Hartree-Fock (HF) calculations without pairing applying the equal filling approximation (EFA) \cite{PerezMartin2008} which have been employed in  Refs.~\cite{Brown2017,Yang2018}.

{\it Correlation analysis}.---Various kinds of correlations between observables and model parameters have been discussed in the context of symmetry energy. The most popular is the  {\it inter-model  analysis}, in which a set of models is used to make predictions and assess systematic uncertainties.   Ideally, the models are supposed to be well calibrated to existing data and sufficiently different in terms of theoretical assumptions and optimization protocols. The implicit assumption here is that
the biases introduced in different models are independent and that theoretical errors are randomized. Examples of such analyses can be found in, e.g.,  Refs.\cite{Brown2000,Roca-Maza2011,Piekarewicz2012,Dutra2012}. The inter-model analysis does not involve any statistical uncertainty quantification; hence, it does not explore  the dependencies between the parameters of the model and the observables studied. 

A more advanced approach to correlations is through  the {\it trend analysis} within a given model. In this case, a selected model parameter or observable $\cal O$, is systematically varied with other  model parameters being calibrated to the optimization dataset. In this way, one can study correlations related to the parameter variations in the direction of {\it one} variable $\cal O$. Studies involving such a trend analysis can be found in, e.g., Refs.~\cite{Klupfel2009,Brown2013,Chen2015}.

In the  {\it statistical correlation analysis}, one probes variations in a full space of model parameters $\vec{p}$. Key quantity is here the penalty function  $\chi^2(\vec{p})$, the root-mean-square deviation of theoretical results from a given 
dataset for the model parameters $\vec{p}$. It structures the space of parameters by the probability $\mathcal{P}(\vec{p})$ for a well tuned set of parameters. The optimal parametrization is found at the minimum of $\chi^2$ where $\mathcal{P}$ has maximum. In its simplest variant, this can be done  by probing the local structure of   $\chi^2(\vec{p})$ surface around its  minimum by studying the Hessian matrix (quadratic approximation) \cite{Reinhard2010,Fattoyev2011} or by means of the Monte Carlo sampling, see, e.g., Ref. \cite{Reinhard2016R}. A more general approach to correlations,  which  probes long-range parameter dependencies,   involves Bayesian calibration. Here, the correlations are directly obtained from posterior distributions of model parameters \cite{Lim2019,Kejzlar2020,Drischler2020a}.

The results of DFT calculations presented in this study are analyzed using the tools
of linear least square regression \cite{Dob14a}. By computing the covariance $\mathrm{cov}(x,y)$ of  quantities $x$ and $y$, as well as their respective variances $\sigma_x$ and $\sigma_y$, we assess statistical
$x$-$y$ correlations  in terms of the bivariate correlation coefficient
\begin{equation}
R_{x,y}=\frac{\mathrm{cov}(x,y)}{\sigma_x\sigma_y}
\end{equation}
or its square $R^2$, which is the coefficient of determination (CoD) \cite{Glantz}. We determine the CoDs as described in Refs.~\cite{Erler2015,Reinhard16}. 
The CoDs contains information on how well one quantity is determined by another one. 

The correlation between two observables can be visualized by the combined probability distribution $P(x,y)$. Each observable which can be described by the given model is also a function of the model parameters, i.e.,  $x=x(\vec{p})$ and y=$y(\vec{p})$. The probability to find certain values $x$ and $y$ is then derived from the probability distribution of $\vec{p}$ as $P(x,y)=\int d\vec{p}(x-x(\vec{p}))(y-y(\vec{p}))\mathcal{P}(\vec{p})$. It becomes a two-dimensional Gaussian distribution in the quadratic approximation mentioned above. That can be well characterized by by equi-probability line $P(x,y)=1/e$ which forms the error ellipsoid in the plane of $x$ and $y$.

Multiple correlation coefficient (MCC) \cite{Allison} of observables
with groups of parameters $\vec{a}$ is:
\begin{equation}
{\rm MCC}(\vec{a},x)=\vec{c}^T(R_{\vecs{a},\,\vecs{a}})^{-1}\vec{c},
\end{equation}
where $R_{\vecs{a},\vecs{a}}$ is the matrix of CoDs between
the model parameters of group $\vec{a}$ and the vector
$\vec{c}=(R_{a_1,x},R_{a_2,x},...)$ contains the
CoDs between the observables and the single group members. Values of
$R^2$ range from 0 to 1, where 0 implies, that those quantities are
completely uncorrelated, 1 denotes that one quantity determines the
other completely.  An $R^2$ of, say,  $0.60$ means that 60\% of the variance in $x$ is predictable from $\vec{a}$, see Refs.~\cite{Schuetrumpf17,Reinhard2018d,Reinhard2020} for recent nuclear physics examples.

\begin{figure}[!htb]
\includegraphics[width=0.8\columnwidth]{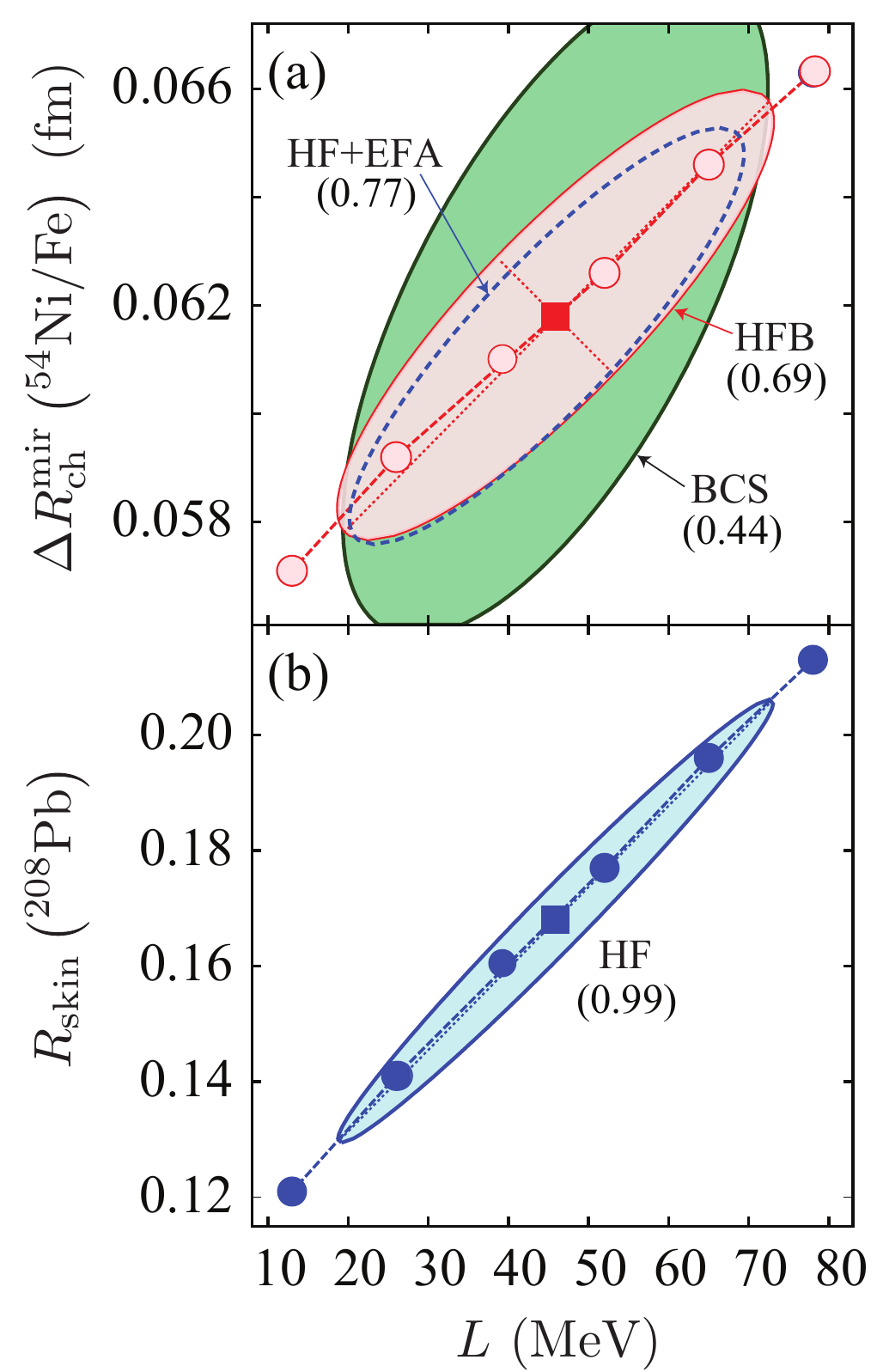}
\caption{\label{fig:mirror-ellipsoids3} The error ellipsoids  in the planes of slope of symmetry energy $L$ and
(a)  mirror radii ${\Rmir} (^{54}$Ni/Fe) and (b) 
$\skin$  in $^{208}$Pb  computed with SV-min. 
the principal axes of the ellipsoids are shown with thin dotted lines, and
the  corresponding  CoDs are indicated by numbers in parenthesis.    The aspect ratio has been chosen such
that a perfectly uncorrelated situation (CoD=0) would
show up as a circle. The results for ${\Rmir} (^{54}$Ni/Fe) were obtained with  HFB, BCS, and HF+EFA. The $\skin$ results for $^{208}$Pb were obtained with HF as the static pairing disappears in this case.
The  circles connected by a
  dotted line show the results from the SV forces with systematically
  varied symmetry energy for the values $L=13,$ 26, 39, 52, 65, and 78 MeV.
  }
\end{figure}
{\it Results}.---To emphasize the need for using HFB in $\Rmir$ calculations, Table\,\ref{Fermilevels} lists the calculated proton Fermi levels for the proton-rich mirror partners considered in this work. These are all very weakly bound ($^{36}$Ca, $^{54}$Ni) or unbound ($^{48}$Ni), which suggests that the proton continuum space can impact theoretical predictions, especially for radial properties \cite{Mil19}.
\begin{table}[htp]
\caption{Proton Fermi levels (in MeV) of  $^{36}$Ca ,  $^{48}$Ni, and   $^{54}$Ni isotopes computed in {\SV} and {\Fy}. }
\begin{ruledtabular}
\begin{tabular}{lcc}
 nucleus & {\SV} & {\Fy} \\
\hline
 $^{36}$Ca & $-$1.0 & $-$0.7 \\
 $^{48}$Ni &   +1.9  &  +3.5 \\
 $^{54}$Ni & $-$2.6  &  $-1.3$ \\
\end{tabular}
\end{ruledtabular}
\label{Fermilevels}
\end{table}

To illustrate the relevance of the CoDs when analysing isovector indicators, Fig.~\ref{fig:mirror-ellipsoids3} compares the  error ellipsoid of SV-min  for ${\Rmir}(^{54}$Ni/Fe) and $L$ with
that for $\skin(^{208}$Pb) and $L$. In the latter case,  the error ellipsoid is very narrow with CoD=0.99, which means that a given value of  $\skin(^{208}$Pb) uniquely determines $L$ within this model. Similarly large is the correlation between $\aD$ in $^{208}$Pb and $L$ with CoD=0.98. The error ellipsoids associated with $\Rmir$ are significantly wider, and they appreciably depend on the treatment of pairing correlations. Namely,  CoD$_{\rm HFB}$=0.69 while CoD$_{\rm BCS}$=0.44. 
The corresponding uncertainties are:
$\sigma_L$(BCS)=26.5\,MeV, $\sigma_L$(HFB)= 27.2 MeV, $\sigma_{\Rmir}$(BCS)=0.0065\,fm, and $\sigma_{\Rmir}$(HFB)=0.0041\,fm. That is,
the uncertainties in $L$ are very similar while $\sigma_{\Rmir}$ strongly depends on the way pairing correlations are treated. This is corroborated by considering also the HF+EFA variant. Here, the correlation (CoD=0.77) is even higher than in the  HFB case. This clearly demonstrates that pairing is responsible for weakening the correlation between $\Rmir$ and $L$.

Figure~\ref{fig:mirror-ellipsoids3} shows also the trends from forces with systematically varied $L$ (dots connected by a line). The trends are nearly linear 
for both $\skin$ and $\Rmir$. However, as illustrated in Fig.~\ref{fig:mirror-ellipsoids3}, the presence of a linear trend is not sufficient to assess the degree of correlation. 
That is, showing that results obtained with different models follow a regular trend is by no means a proof of a strong correlation. Indeed  a CoD=0.44  ($R=0.66$) implies a  moderate correlation while CoD=0.69  ($R=0.83$) is indicative of a stronger correlation. It is only in the case of $\skin$ in which the major axis of the correlation ellipsoid practically coincides with the systematic trend line that a near-perfect correlation is obtained that can be used for the parameter reduction. It is interesting to note  
the large  CoD values around 0.86-0.88 ($R_{L,\Rmir}$=0.93-0.94) that were obtained 
for $\Rmir$ in $^{54}$Ni/Fe and other mirror pairs 
in Ref.~\cite{Yang2018}. In our opinion, the reason for the large correlations obtained in Ref.~\cite{Yang2018} is twofold. First, they ignore pairing correlations. Second, their relativistic energy density functionals have only two isovector parameters (instead of six for Skyrme functionals), which automatically implies strong correlations between all isovector observables \cite{Reinhard2010}.

\begin{figure}[!htb]
\includegraphics[width=1.0\columnwidth]{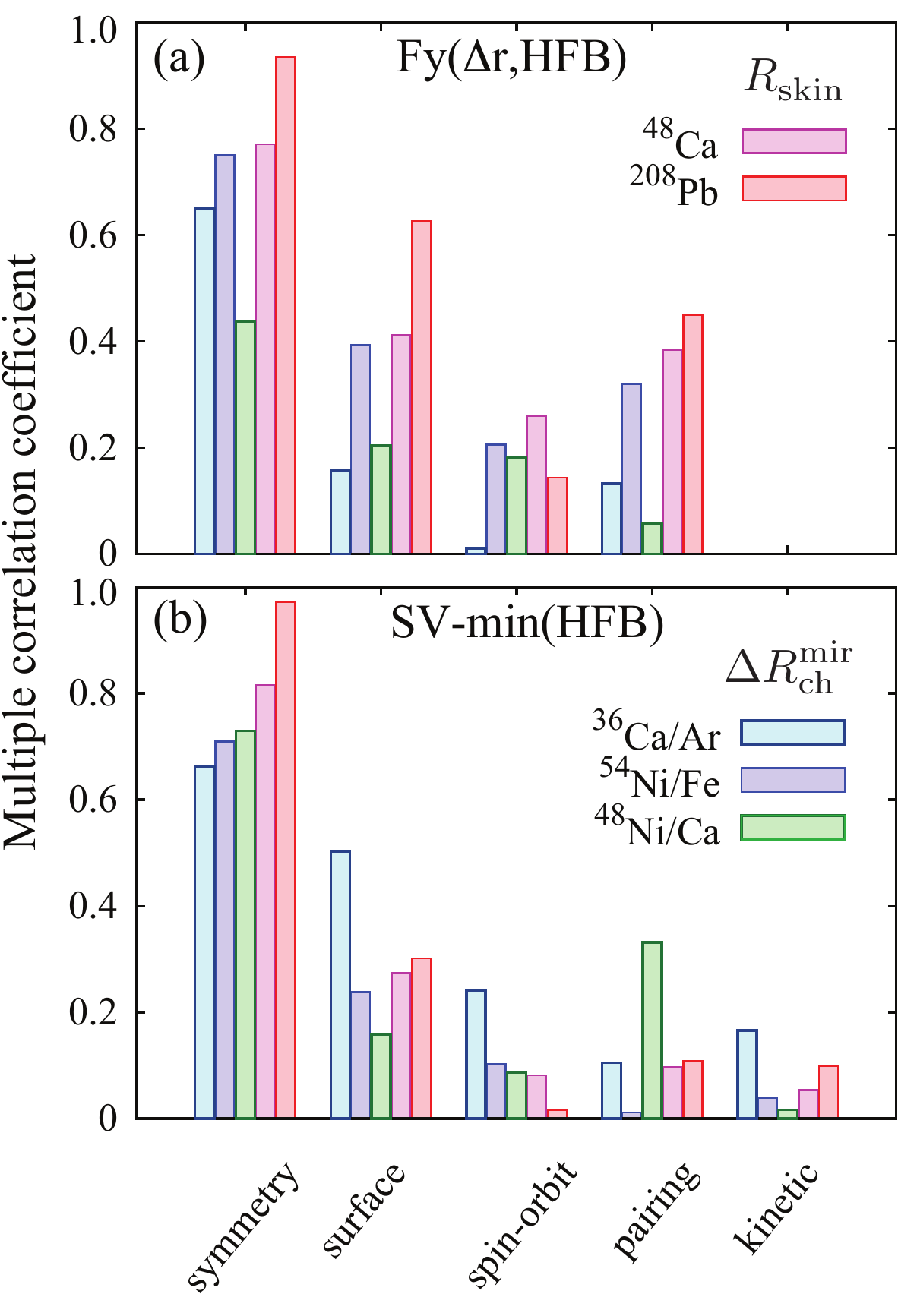}
\caption{\label{fig:MCC-mirror-skin} The multiple correlation
  coefficients  between  various observables ($\skin$ and $\Rmir$ in different nuclei) and
   groups of selected model parameters for SV-min(HFB) and \Fy:  symmetry energy
  parameters $J$ and $L$, isoscalar and isovector surface energy
  parameters, isoscalar and isovector spin-orbit  parameter,  pairing parameters, and the isoscalar and isovector effective
  masses as parameters of the kinetic energy.}
\end{figure}

\begin{figure}[!htb]
\includegraphics[width=0.8\columnwidth]{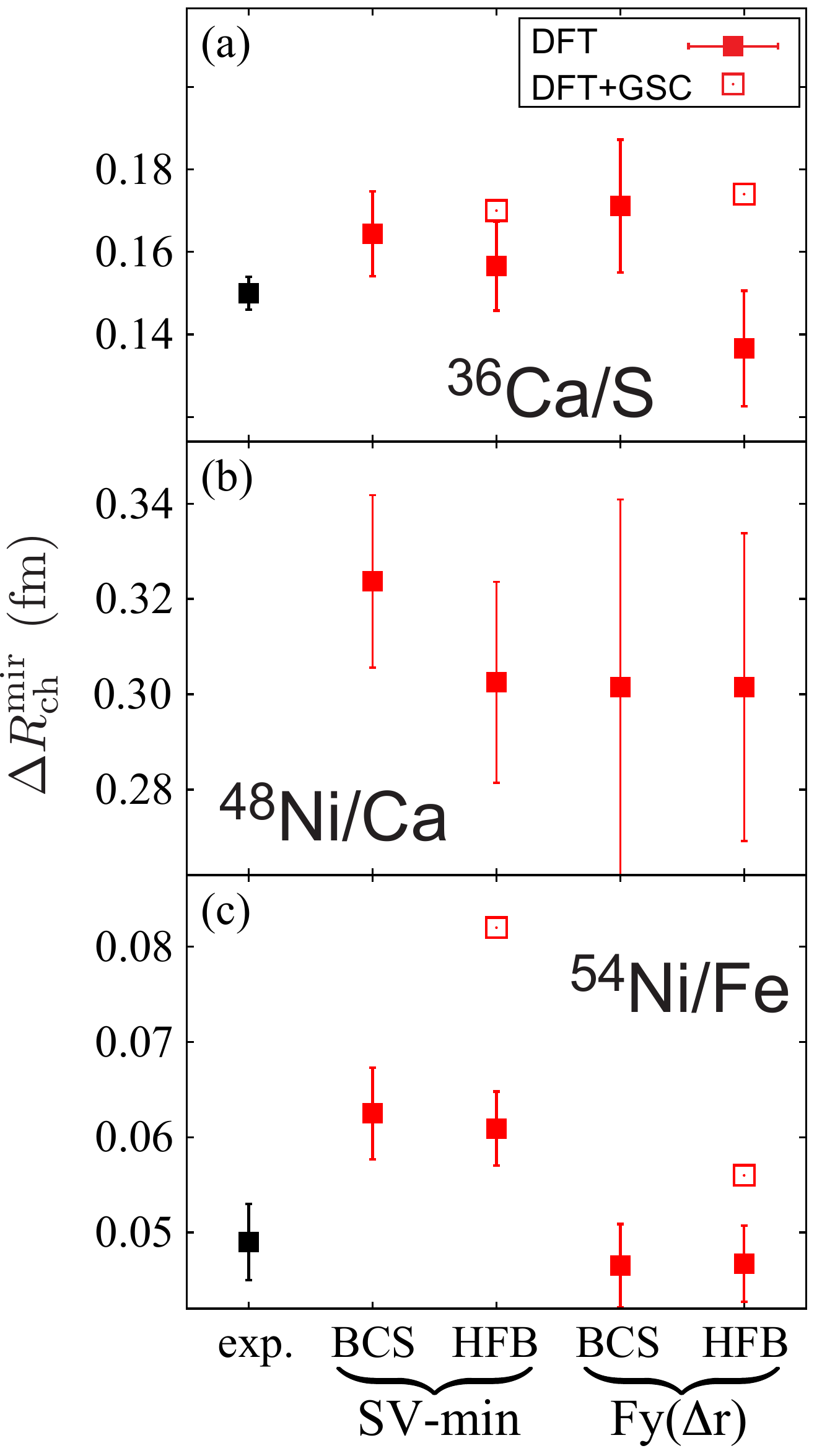}
\caption{\label{fig:diffradii2} The mirror radius differences
for (a) $^{36}$Ca/S, (b) $^{48}$Ni/Ca, and (c) $^{54}$Ni/Fe
calculated with SV-min and FY($\Delta r$) EDFs  in BCS and HFB variants. 
 The
  error bars indicate the statistical uncertainty associated with
$\chi^2$ optimization. For orientation, results HFB calculations including collective ground state correlations in  $^{36}$Ca/S  and  $^{54}$Ni/Fe are shown by open squares.  The experimental $\Rmir$ values are   shown for
$^{36}$Ca/S \cite{Brown2020} and  $^{54}$Ni/Fe \cite{Pineda2021}  (black squares). 
  }
\end{figure}

The fact that mirror radii differences  produce  broader error ellipsoids as compared to those of  excellent isovector indicators such as  $\skin$, suggests that they are influenced also by other terms in the EDF parametrization as, e.g., surface energy. This is analyzed systematically in Fig.\,\ref{fig:MCC-mirror-skin} which shows the MCCs between $\skin$ and $\Rmir$ in different nuclei and groups of selected parameters characterizing SV-min(HFB) and {\Fy}.
As expected, $\skin$ in $^{208}$Pb is practically determined by the symmetry energy parameters. This correlation is somehow reduced for $\skin$ in a lighter nucleus  $^{48}$Ca, because it is more impacted by shell effects.
The symmetry energy still dominates MCCs with mirror radii differences, but its impact is not as pronounced as in the case of  $\skin$ in $^{208}$Pb. The values of $\Rmir$  are in fact influenced by many terms of the functional, i.e., they are {\it distributed quantities} \cite{Schuetrumpf17,Reinhard2018d}, which is indicative of shell effects. The corresponding MCCs
still deliver useful information, but only when combined with other data in a consistent statistical analysis.

The  values of $\Rmir$ for $^{36}$Ca/S, $^{48}$Ni/Ca, and  $^{54}$Ni/Fe mirror pairs calculated in BCS and HFB   are shown in Fig.~\ref{fig:diffradii2}. In general, there is a good agreement with experiment considering the fact that the expected accuracy for precision calculation of charge radii is 0.015\,fm \cite{Reinhard2021so}. The treatment of pairing does  affect 
$\Rmir$, especially for weakly-bound $^{36}$Ca and unbound $^{48}$Ni, see Table~\ref{Fermilevels} and Ref.~\cite{Mil19}. 
To get some idea on the impact of zero-point correlations on mirror radii,  we also show in Fig.~\ref{fig:diffradii2}  the effect of collective ground state correlations on charge radii from low lying
$2^+$ states  \cite{Kluepfel2008}. Those beyond-DFT corrections 
are quite significant for $\Rmir$($^{54}$Ni/Fe). In this context, we note that the observed $B(E2)$ rates in mirror nuclei indicate the presence of strong isovector effects  \cite{Wimmer2021}, which are likely to affect $\Rmir$.

\begin{figure}[!htb]
\includegraphics[width=\columnwidth]{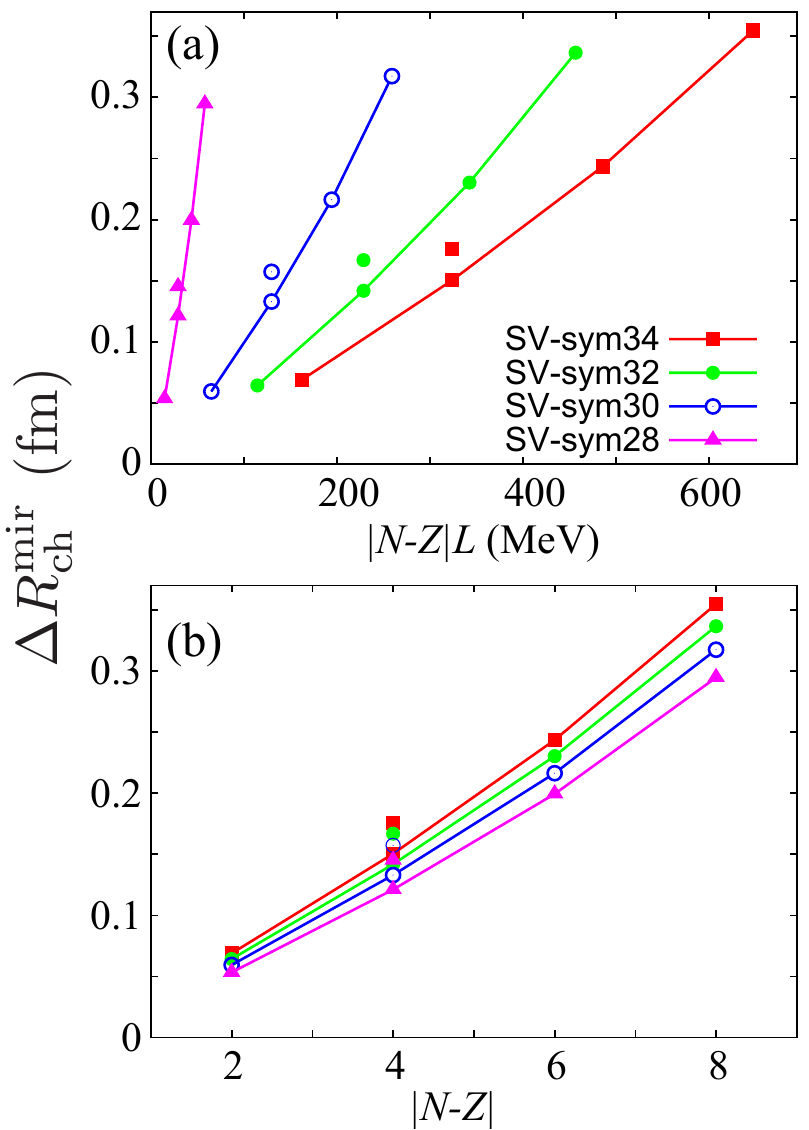}
\caption{\label{fig:systematics}
The mirror radius differences for
$^{54}$Ni/Fe, $^{52}$Cr/Ni, $^{50}$Ti/Ni, and $^{48}$Ca/Ni mirror pairs, and for the single mirror pair $^{36}$Ca/S  as functions of (a) $|N-Z|L$ and (b) $|N-Z|$ calculated for four SV EDFs with constrained symmetry energy $J$.}
\end{figure}
It has been suggested in Ref.\,\cite{Brown2017} that differences in mirror charge radii $\Rmir$ are proportional to $|N-Z|L$. As seen in Fig.~\ref{fig:systematics}(a), this does not seem to be the case for the SV-family of EDFs with systematically varied symmetry energy. On the other hand, the values of $\Rmir$ seem to scale with $|N-Z|$ \cite{Gaidarov2020} or $|N-Z|/A$ \cite{Novario2021} (we checked that the scaling  with  $|N-Z|/A$ produces very similar results).

{\it Conclusions}.---In this study, we examine claims
\cite{Brown2017,Yang2018} that the precise data on mirror charge radii can provide a stringent constraint on $L$. While our statistical analysis confirms an appreciable correlation between $L$ and $\Rmir$, this correlation is significantly weaker than that between $L$, $\skin$, and $\aD$  in $^{208}$Pb. In other words, we find $\Rmir$ to be a much weaker isovector indicator than neutron skins in heavy nuclei,  dipole polarizability, or  parity-violating asymmetry
\cite{ReinhardPREX}.

Since the proton-rich nucleus in the mirror pair is usually weakly bound, special care should be taken when considering the impact of the proton continuum space. This can be done, e.g., by employing the HFB formalism.  

Pairing correlations should always  be considered for open-shell  nuclei. As shown in our paper,
by neglecting pairing~\cite{Brown2017,Yang2018,Brown2020,Pineda2021}  one artificially increases the  correlation between $\Rmir$  and $L$.

The statistical errors on predicted $\Rmir$ vary between 0.005-0.02\,fm, depending on $|N-Z|$. The systematic uncertainties, however, are significantly larger. Those are due to the choice of nuclear interaction, treatment of pairing, and evaluation of zero-point collective correlations in  spherical and transitional nuclei.

Considering the  large theoretical uncertainties on $\Rmir$ estimated in our study, we conclude that the precise data on mirror charge radii
with an error of about 0.005\,fm \cite{Brown2017}, while extremely valuable for studying isospin effects in nuclei and model developments, {\it cannot} provide a stringent constraint on $L$.

{\it Acknowledgements}.---This material is based upon work supported by the U.S.\ Department of Energy, Office of Science, Office of Nuclear Physics under award numbers DE-SC0013365 and DE-SC0018083 (NUCLEI SciDAC-4 collaboration).
We also thank the RRZE computing center of the Friedrich-Alexander university Erlangen/N\"urnberg for supplying resources for that work.

\bibliography{references}
\end{document}